\documentclass[aps,prd,amsmath,amssymb,showkeys,showpacs]{revtex4}
\usepackage{graphicx}% Include figure files
\usepackage{bm}% bold math\usepackage{graphicx}% Include figure files
\newcommand{\be}{\begin{equation}}
\newcommand{\ee}{\end{equation}}
\newcommand{\bea}{\begin{eqnarray}}
\newcommand{\eea}{\end{eqnarray}}
\newcommand{\hf}{\frac12}
\newcommand{\nn}{\nonumber\\}
\def\eq#1{(\ref{#1})}
\def\journal#1#2#3#4{{#1} {\bf #2}, #3 (#4)}
\def\la{\langle}
\def\ra{\rangle}
\def\Tr{{\mathrm{Tr}}}

\def\mr#1{{\mathrm{#1}}}
\def\v#1{{\bm{#1}}}
\def\ord#1{{\cal O}(#1)}

\def\fdd#1#2#3{\frac{\delta^2#1}{\delta#2\delta#3}}

\def\hj{{\hat j}}
\def\hsi{\hat\psi}
\def\hsid{\bar\psi^\dagger}

\def\hu{{\hat u}}

\def\hD{{\hat D}}
\def\hG{{\hat G}}
\def\ih{\frac{i}{\hbar}}
\def\psid{\psi^\dagger}
\def\tib{$T\hskip-6pt/$}

\begin{document}
\title{The semiclassical Coulomb field}
\author{J. Polonyi}
\homepage{http://lpt1.u-strasbg.fr/polonyi/homepage.html}
\affiliation{Theoretical Physics Laboratory, Louis Pasteur University, Strasbourg, France}
\date{\today}
\begin{abstract}
The contribution of different modes of the Coulomb field to decoherence and to the dynamical
breakdown of the time reversal invariance is calculated in the one-loop approximation
for non-relativistic electron gas. The dominant contribution was found to come from the usual 
collective modes in the plasma, namely the zero-sound and the plasmon oscillations. The length scale of 
the quantum-classical transition is found to be close to the Thomas-Fermi screening length. It is 
argued that the extension of these modes to the whole Fock-space yields optimal pointer states.
\end{abstract}
\maketitle

\section{Introduction}
Any attempt to bridge the gap between quantum and classical physics must be based on the treatment
of a large number of degrees of freedom. This requires the use of the formalism of Quantum Field Theory.
The systematical development in this domain started with the introduction of the interaction representation 
\cite{tom,schwe} to obtain the perturbation series for expectation values. Later, being interested mainly in 
cross sections, the calculations were rendered simpler by the introduction of the scattering matrix formalism. 
But this restricted formalism which received further, strong support form the path integral representation of 
the transition amplitudes is not well suited for the description of quantum-classical transition where reduced 
density matrices should be studied. Instead, the return to the more involved Heisenberg or interaction 
representation gives us access to the density matrix of the system and opens the way to develop
theoretical methods to approach the quantum-classical transition. The main point of this work is to 
show in the framework of non-relativistic electron gas that Schwinger's closed time path (CTP) 
formalism \cite{schw,keldysh} is well suited to these goals.

Let us consider an observable $A$ with expectation value 
\be\label{expv}
\la\psi(t)|A|\psi(t)\ra=\Tr[A\rho(t)]
\ee
where the density matrix at time $t$ is given by
\be
\rho(t)=e^{-\ih(t-t_i)H}\rho_ie^{\ih(t-t_i)H}
\ee
in terms of the initial density matrix $\rho_i$ given at the initial time $t_i$. Such an expectation value in 
the Heisenberg representation can not, in general, be reproduced by the usual vacuum-to-vacuum transition 
amplitudes, such as
\be\label{trampl}
\la\psi(t_i)|e^{-\ih(t_f-t)H}Ae^{-\ih(t-t_i)H}|\psi(t_i)\ra,
\ee
 used in constructing the scattering amplitudes. In fact, the expectation value \eq{expv}, obtained for
$\rho_i=|\psi(t_i)\ra\la\psi(t_i)|$ is related to \eq{trampl} when $|\psi_i\ra$ is an eigenstate of the 
full Hamiltonian, a rather trivial and unrealistic case. One can bring over the techniques and the experience
obtained in the scattering matrix formalism to the CTP formalism which aims at the expectation
values \eq{expv} rather than the transition amplitudes \eq{trampl}.

There are two time axes appearing in operators of the Heisenberg representation
which lead to a formal reduplication of the degrees of freedom in calculating the
expectation values. This has the following two important features. One is the 
identification of the combinations of the
degrees of freedom which couple to retarded or advanced Green functions \cite{ed}, giving
a direct way to access the possible breakdown of the time reversal invariance. The other is that
this reduplication leads us to the density matrix rather than the transition amplitude.

There are two known signatures of the quantum-classical transition. One is the emergence of
classical probabilities, called decoherence \cite{zeh,zurek}.
Let us use the basis $|S_n\ra$ for the system plus measuring apparatus, denote the environment
state vector by $|E\ra$ and suppose that the whole system starts with the initial
state $\sum_nc_n(t_i)|S_n\ra\otimes|E_0\ra$ at  $t=t_i$ which evolves into 
$\sum_nc_n(t)|S_n\ra\otimes|E_n\ra$ for $t>t_i$ due to the unavoidable interactions between the
system and the environment. Actually, the features 'macroscopic' and the 'impossibility of separating
from the environment' are believed to be equivalent. The interactions with the environment are always 
strong because the latter is macroscopic and its gap-less, dense energy spectrum makes easy the jump $|E\ra\to|E'\ra$
with $\la E|E'\ra=0$ even by a small amount of energy exchange 
with the system $S$. As a result, one expects $\la E_n|E_m\ra\approx0$ for $n\not=m$ and
the reduced density matrix for the system, 
\bea
\rho_S&=&\Tr_E\sum_{n,n'}c^*_nc_{n'}|S_n\ra\otimes|E_n\ra\la E_{n'}|\otimes\la S_{n'}|\nn
&\approx&\sum_n|c_n|^2|S_n\ra\la S_n|
\eea
is approximately diagonal, indicating the suppression of interference between system states
which correspond to macroscopically different environment states. 
The reduced density matrix, being hermitian, is always diagonal when expressed in an appropriately chosen
basis. The observables which are diagonal in this basis are called pointers \cite{zurek}. 
Decoherence makes the pointer probabilities $p_n=P(S=S_n)$ additive,
$P(S=S_n\cup S=S_{n'})=P(S=S_n)+P(S=S_{n'})$ for $n\not=n'$  as in the usual probability theory.

The other signature we have to establish at the quantum-classical transition is more involved, it is
the dynamical breakdown of the time reversal invariance, \tib. This is needed to read off the result of the 
measurement, ie. to create durable, macroscopic records which continues to exists independently of the
system with its microscopic, time reversal invariant dynamics.

Both signatures can be realized in the limit of large number of degrees of freedom only. 
The suppression of the overlap of two states in decoherence is achieved by the multiplication
of a large number of overlap factor, each of them being between zero and one in absolute magnitude,
belonging to the microscopic degrees of freedom of the environment. The dynamical symmetry
breaking, \tib, requires degeneracy of the non-interacting system-environment states,
to 'leak' the former into the latter. Such a high degree of degeneracy can be achieved in 
the thermodynamical limit only. 
We shall see that these two mechanisms driving the quantum-classical transition
become identical in the CTP formalism. This is due to their common dynamical origin, the 
continuous, gap-less spectrum and the resulting approximate degeneracy of the environment states.

The thermodynamical limit implies another, related dynamical process, the screening. It is the
result of the presence of soft excitation modes which can efficiently polarize the vacuum.
Does the screening has a role to play in the quantum-classical transition? The reason
of suspecting this is that the dressed quasi-particles obtained by vacuum polarization
have extended structure and may represent the pointer states which are stable and
decohere easily \cite{zeh,zurek,pointer}. It will be shown that the CTP formalism, applied to the non-relativistic 
electron gas provides an affirmative answer to our question. It is found that the zero-sound and the plasmon 
collective modes are responsible for decoherence and \tib~in the electron plasma and these processes
appear just at the screening length scale. This is a non-trivial result because the screening is realized
in a static system while the other two mechanisms are based on time-dependent modes.

The quantum-classical transition is considered in this work from the point of view of the renormalization group,
when the modes are eliminated from the dynamics starting at short distances and going towards the
infrared limit. This process is regarded on a rough, qualitative level and our goal is only to characterize
the contributions of the eliminated modes from the point of view of their contributions to classical behavior.
For this end we introduce the classicality of a quasi-particle mode characterized by a wave vector and frequency and 
calculate it for homogeneous, non-relativistic electron plasma in the one-loop approximation. 

The concept of classicality has already been introduced in a number of works. It has been used to 
characterize states form the point of view of the loss of information \cite{optpoint}, the robustness \cite{wiva}, 
the impact on the environment \cite{dzdazu}, and has been compared in \cite{dadzzu}. The loss of information
and stability were compared in \cite{elze}. An intuitive proposition, based
on the view of measurement as fluctuation induced macroscopic instability is outlined in \cite{dreyer}.
The classical properties of states in the Hilbert space has been considered in \cite{grkemo}.
Another state-dependent definitions, based on the phase space distribution functions were presented 
in \cite{dore,mapa}. Our definition of classicality, whose inverse can be considered as a measure of the 
distance from the classical domain, provides a more detailed picture of the establishment of classical
physics because it is applicable for quantum modes, characterized by energy-momentum
rather than states as in the previous examples.

The organization of the paper is the following. The effective CTP theory for the Coulomb field is derived in 
the one-loop approximation in Section \ref{effacf}. Section \ref{classs} contains the identification of
the contribution of each mode of the Coulomb field to the reduced density matrix and the  
introduction of classicality, a measure of the strength of modes to generate decoherence and \tib. 
This quantity is computed for the electron plasma and followed during the gradual decrease of the infrared
cut-off in space or time in Section \ref{points}.  It is found that the zero-sound and plasmons are the modes 
which drive the transition to classical behavior. Finally, Section \ref{concls} is reserved for the conclusions.
There are two Appendices to summarize some technical details. Appendix \ref{ctpprop}
contains the brief derivation of the CTP propagator for non-relativistic particles at finite
density and temperature. The calculation of the one-loop self energy of the Coulomb field, the non-trivial
dynamical input to our effective theory is sketched in Appendix \ref{tgap}.

\section{Effective theory for the Coulomb field}\label{effacf}
Our goal is the construction of  the reduced density matrix for the Coulomb field $A_0(x)=u_x$,
$x=(t,\v{x})$, in the presence of non-relativistic electrons of finite, homogeneous density. 
We start with the generator functional for the insertion of the Coulomb field into the evolution of the initial density matrix,
\be
e^{\ih W[j^+,j^-]}=\Tr T[e^{-\ih\int_{t_i}^{t_f}dt'[H(t')-\int d^3xj^+(t,\v{x})u(t,\v{x})]}]\rho_i
\bar T[e^{\ih\int_{t_i}^{t_f}dt[H+\int d^3xj^-(t,\v{x})u(t,\v{x})]}]
\ee
where $H$ denotes the Hamiltonian of the electrons interacting with the Coulomb potential
and $\bar T$ stands for anti-time ordering. Each degree of freedom of the system follows unconstrained 
time evolution and we retain no information from their future behavior in defining the Green functions 
for the Coulomb field. Thus the trace extends over the whole Fock-space. 

The path integral representation of the generating functional is 
\be\label{pintgf}
e^{\ih W[\hj]}=\int D[\hu]D[\hsid]D[\hsi]e^{\ih\hsid\cdot[\hG^{-1}-e\hat\sigma\hu]\cdot\hsi
+\frac{i}{2\hbar}\hu\cdot\hD_0^{-1}\cdot\hu-\frac{ie}{\hbar}\hu\cdot\hat n+\ih\hj\cdot\hu}
\ee
where the hat denotes a CTP doublet, $\hu=(u^+,u^-)$, etc. and
\be
\hat\sigma=\begin{pmatrix}1&0\cr0&-1\end{pmatrix}.
\ee
As usual, the initial state can be chosen in an arbitrary manner as long as it has an overlap with the 
true vacuum on the expense of carrying out the limit $t_i\to-\infty$. In particular, $|\psi(t_i)\ra$ is a non-interacting 
Fermi-sphere. The initial wave functional for the Coulomb field is chosen to be a constant. The final time boundary conditions
are $u^+_{t_f,\v{x}}=u^-_{t_f\v{x}}$ and 
$\psi^+_{t_f,\v{x}}=\psi^-_{t_f,\v{x}}$, $\psi^{+\dagger}_{t_f,\v{x}}=\psi^{-\dagger}_{t_f,\v{x}}$.
The space and space-time integrations, together with the summation over spin index if necessary, are indicated 
by a scalar product ie. $f\cdot g$ stands for $\int dxf_xg_x$ or $\int d^3xf_\v{x}g_\v{x}$ and
$\psid\cdot\psi=\sum_\sigma\int dx\psid_{\sigma x}\psi_{\sigma x}$. The inverse propagators
$\hG^{-1}=\hG^{-1}_0+\hG^{-1}_\mr{BC}$ and $\hD^{-1}_0=\hD^{-1}_B+\hD^{-1}_\mr{BC}$ contain 
the usual terms of the single time axis formalism,
\bea
\hG_0^{-1}&=&\begin{pmatrix}G_0^{-1}&0\cr0&-G_0^{-1*}\end{pmatrix},\nn
\hD^{-1}_B&=&\begin{pmatrix}D_0^{-1}&0\cr0&-D_0^{-1*}\end{pmatrix},
\eea
with
\be
G^{-1}_{0~(\sigma x),(\sigma' x')}=\delta_{\sigma\sigma'}
\left(i\hbar\partial_t+\frac{\hbar^2}{2m}\Delta_x+\hbar\mu+i\epsilon\right)\delta_{x,x'}
\ee
and
\be\label{phtpr}
D_{0~xx'}=(-\Delta_x+i\epsilon)\delta_{x,x'}.
\ee
The boundary condition term, $\hG^{-1}_\mr{BC}$ implements the boundary conditions in time.
 Finally, the homogeneous particle density 
$\hat n=(n,-n)$ is introduced to neutralize the system and thereby remove the tadpole contributions.

The Coulomb field connected Green functions, monitored by their generating functional $W[\hj]$, characterize
the subsystem of the Coulomb photons in the environment provided by the electron gas. In particular, the
effective theory obtained by eliminating the latter leads to the dynamics of the Coulomb photons. 
The integration over the fermion field yields the bosonized path integral
\be
e^{\ih W[\hj]}=\int D[\hu]e^{\Tr\ln[\hG^{-1}-e\hat\sigma\hu]+\frac{i}{2\hbar}\hu\cdot\hD_0^{-1}\cdot\hu
-\frac{ie}{\hbar}\hu\cdot\hat n+\ih\hj\cdot\hu}
\ee
over the Coulomb field alone. The expansion
\be
\ln[\hG^{-1}-e\hat\sigma\hu]=\ln\hG^{-1}-\sum_{n=1}^\infty\frac{1}{n}\Tr(e\hG\cdot\hat\sigma\hu)^n
\ee
leads to the effective theory for the Coulomb field,
\be\label{efth}
e^{\ih W[\hj]}=\int D[\hu]e^{\ih S_\mr{eff}[\hu]+\ih\hj\cdot\hu},
\ee
governed by the effective action
\be\label{ceffa}
S_\mr{eff}[\hu]=-i\hbar\Tr\ln\hG^{-1}+\hf\hu\cdot\hD^{-1}\cdot\hu-e\hat k\cdot\hu+\ord{\hu^3},
\ee
involving the charge density
\be
\hat k_x=\hat n+i\hbar(\hG\hat\sigma)_{x-\eta e^0,x+\eta e^0},
\ee
given in terms of the CTP propagator for the electron field, $\hG$, cf. Appendix \ref{ctpprop}.
The neutralizing background charge density is chosen in such a 
manner that $\hat k$ is vanishing. The inverse of the dressed Coulomb propagator,
\be
\hD^{-1}=\hD_0^{-1}-\hat\Sigma,
\ee
is expressed by means of the self energy which is
\be\label{cself}
\hat\Sigma^{\sigma\sigma'}_{xx'}=-ie^2\hbar\sigma\sigma'G^{\sigma\sigma'}_{xx'}G^{\sigma'\sigma}_{x'x}
\ee
in the one-loop approximation.

The Green functions encapsulated in the functional $W[\hj]$ represent the moments of the reduced density matrix
for the Coulomb field. The quadratic approximation to the effective action for the Coulomb field shows clearly
the modes which build up these moments. The one-loop result, summarized briefly in Appendix \ref{tgap}, yields
\be\label{hdinv}
\hD^{-1}_{\omega,\v{q}}=\begin{pmatrix}\v{q}^2-L_{\omega,\v{q}}+i\epsilon&0\cr0&-\v{q}^2+L_{\omega,\v{q}}+i\epsilon\end{pmatrix}
+ir_{|\omega|,\v{q}}\begin{pmatrix}1&-2\Theta(-\omega)\cr-2\Theta(\omega)&1\end{pmatrix}+\hD_\mr{BC}^{-1},
\ee
in Fourier space where 
\be\label{lindhfn}
L_{\omega,\v{q}}=\frac{e^2mk_F}{2\pi^2\hbar^2}\left\{-1+\frac{1}{2q}\left[1-\left(\frac{z}{q}-\frac{q}{2}\right)^2
\right]\ln\left|\frac{1+\frac{z}{q}-\frac{q}{2}}{1-\frac{z}{q}+\frac{q}{2}}\right|
-\frac{1}{2q}\left[1-\left(\frac{z}{q}+\frac{q}{2}\right)^2
\right]\ln\left|\frac{1+\frac{z}{q}+\frac{q}{2}}{1-\frac{z}{q}-\frac{q}{2}}\right|\right\}
\ee
with $z=\omega m/\hbar k_F^2$ and $q=|\v{q}|/k_F$ and
\be\label{limag}
r_{\omega,\v{q}}=\Theta(z)\frac{me^2k_F}{4\pi\hbar^2q}\begin{cases}1-\left(\frac{z}{q}-\frac{q}{2}\right)^2&
q>2,~~~\frac{q^2}{2}-q<z<\frac{q^2}{2}+q,\cr
1-\left(\frac{z}{q}-\frac{q}{2}\right)^2&q<2,~~~q-\frac{q^2}{2}<z<\frac{q^2}{2}+q,\cr
2z&q<2,~~~0<z<q-\frac{q^2}{2}.\end{cases}
\ee

We parametrize the density dependence by means of the usual dimensionless parameter $r_s=r_0/a_0$
denoting the average electron-electron separation, $r_0=(3V/4\pi N)^{1/3}$, in units of the Bohr radius, $a_0$.
The relation $e^2m/\hbar^2k_F=4\pi\kappa r_s$, $\kappa=(4/9\pi)^{1/3}$ can be used to arrive at the expressions
\be
\v{q}^2-L=k_F^2\left[q^2-\frac{2\kappa r_s}{\pi}\left\{\frac{1}{2q}\left[1-\left(\frac{z}{q}-\frac{q}{2}\right)^2\right]
\ln\left|\frac{1+\frac{z}{q}-\frac{q}{2}}{1-\frac{z}{q}+\frac{q}{2}}\right|
-\frac{1}{2q}\left[1-\left(\frac{z}{q}+\frac{q}{2}\right)^2\right]
\ln\left|\frac{1+\frac{z}{q}+\frac{q}{2}}{1-\frac{z}{q}-\frac{q}{2}}\right|-1\right\}\right]
\ee
and
\be
r_{\omega,\v{q}}=\Theta(z)k_F^2\frac{\kappa r_s}{q}\begin{cases}
1-\left(\frac{z}{q}-\frac{q}{2}\right)^2&q>2,~~~\frac{Q^2}{2}-q<z<\frac{q^2}{2}+q,\cr
1-\left(\frac{z}{q}-\frac{q}{2}\right)^2&q<2,~~~q-\frac{q^2}{2}<z<\frac{q^2}{2}+q,\cr
2z&q<2,~~~\frac{q^2}{2}-q<z<q-\frac{q^2}{2}.\end{cases}
\ee

The physical interpretation of the effective action \eq{ceffa} is the following. 
The real part of the diagonal components in \eq{hdinv}, 
$\Re(\hD^{-1})^{++}$, describes screening and determines the space-time structure of the quasi-particles, 
defined by the dispersion relation $\Re(\hD^{-1})^{++}_{\omega,\v{q}}=0$. The imaginary part of the 
diagonal component, $\Im(\hD^{-1})^{++}$, reflects Landau-damping which is the leakage of the Coulomb photons 
into electron-hole pairs and generates finite life-time for the dressed Coulomb photons. 

The non-diagonal matrix elements in Eq. \eq{hdinv} display entirely different physical processes. 
Notice that the variables residing on different time axes in the original, microscopic description \eq{pintgf} 
are coupled at the final time by the boundary conditions only. The couplings between the time axes at intermediate 
time contribute to non-factorisable terms in the density matrix and stand for entanglement in the system, 
that of the Coulomb photons interacting with the electron gas in our case. This contribution is purely imaginary
in the Fourier-space, the suppression produced by it is just decoherence, cf. \eq{effevfsp} below
which indicates that the one-loop entanglement is decoherence. The structure of the CTP self energy, the
common factor appearing in the diagonal and off-diagonal imaginary matrix element in Eq. \eq{hdinv},  
shows the identical dynamical origin of the finiteness of the 
life-time and decoherence as far as the one-loop quasi-particle structure is concerned.

The effective action \eq{ceffa} is actually the influence functional \cite{feynman} for the Coulomb field and the
contributions generated by the elimination of the electrons, the self-energy in our approximation, are
independent on the boundary conditions imposed on the photon field. By opening the final boundary
conditions for the photons the dependence on the final configurations of the generating functional 
reproduces the reduced density matrix.

\section{Classicality}\label{classs}
The transition from the quantum to the classical regime is supposed to be driven by the decoherence and the 
dynamical breakdown of the time reversal invariance. The interaction Lagrangian density,
$-e\psid_x\psi_xu_x$, is given in terms of the fields thus one expects that the suppression of the 
off-diagonal matrix elements of the reduced density matrix will be achieved in the field diagonal basis. 

The other ingredient of the classical limit, \tib, is to establish durable records of macroscopic events. 
One should separate two obvious aspects from the non-trivial, dynamical aspects of \tib. 
The first trivial feature is the appearance of the retarded Lienard-Wiecher
potentials in classical electrodynamics even though the Maxwell equations are invariant under time
reversal. In fact, it is the Cauchy problem with well defined initial conditions which makes the
retarded Green functions to appear in the classical equations of motion. Had the final conditions been given to render 
the time evolution well defined, the advanced Lienard-Wiecher potential should have been used. The nontrivial 
aspect of \tib~in classical physics is the mixing \cite{arnold}, the gradual and irrecoverable spread of informations 
in non-integrable systems. Such a dynamics generates a well defined time arrow for observers equipped with finite 
resources or resolution.  The information represented by the initial conditions is spread in this manner
in chaotic, classical dynamical systems. There is no mixing in quantum dynamics
due to the linearity of the Schr\"odinger equation but the increased sensitivity to the initial
conditions for long time evolution of classically chaotic systems \cite{peres} still shows 
some similarity with the mixing of classical physics. 

The second triviality is that the finite life-time is not yet \tib. The induced dynamics of a subsystem is unitary, the
scalar product of states of the subsystem is obviously preserved in time. It is easy to see that another, necessary 
condition for \tib~beside the finite life-time is that the environment have no gap in its excitation spectrum. Let us consider the
natural line width, as an example. It generates \tib~because there are infinitely many soft photons
wound up around an asymptotic electron state and they can not be resolved with finite resources. 
It is instructive to see briefly what happens in the superconducting vacuum. This latter is a macroscopic quantum
effect in the absence of \tib~, the life-time of excited atomic states remain finite, their decay can be resolved and
the exclusive cross sections are finite because finite number of photons participating in the dressing.

Let us now turn to the inverse problem, the effective dynamics of photons in the presence of electrons.
The opening of channels above the pair creation threshold where photons decay into real electron-positron 
pairs is indicated by the non-vanishing imaginary part of the photon self energy, the generation of finite life-time 
for sufficiently energetic photon states. The environment of the photons, the Dirac-see has a gap and finite number
of on-shell electron-positron pairs are created only. The on-shell nature of the created particles is important because it 
assures that the energy released by the pair creation process will be diffused. In order to lower the pair creation threshold 
we consider QED in the presence of a non-vanishing charge density, realized by the introduction of a chemical 
potential which places the Fermi level into the positive energy continuum of the one-particle spectrum.
An infinitesimal energy is now sufficient for the polarization of the Fermi sphere, the
creation of particle-hole pairs. A photon can decay into infinitely many on-shell, propagating particle-hole
pairs, $|\gamma\ra\to|\gamma\ra+|\gamma,p,h\ra+|\gamma,2p,2h\ra+\cdots$
where $p$ and $h$ denote particle and hole excitations. Such a spread of the 
one-photon state into other propagating states of the Fock-space represents a diffusion of the information
and appears as the quantum analogy of mixing. 

It is the restriction of our description into a subsystem in a gap-less, 'diffusive', environment which generates 
non-unitary time evolution. In other words,  \tib~appears as soon as the finite resolution power of
observations leads to loss of informations. Observations which can resolve all particle-hole pairs 
surrounding a photon find no violation of the time reversal invariance and could isolate any component of the 
dressed photon state. 

Let us see now our two signatures of the quantum-classical crossover. To identify first the decoherence mechanism 
we take the plane wave
\be
u^\pm_{t,\v{x}}=u^\pm_{\omega,\v{q}}\cos(\omega t-\v{q}\v{x})
\ee
representing the one-loop dressed quasi-particles and use the parametrization 
$u^\pm=u\pm v/2$ which yields the effective action
\be\label{effevfsp}
S_\mr{eff}(u,v)=TV\left[u_{\omega,\v{q}}v_{\omega,\v{q}}(\v{q}^2-L_{\omega,\v{q}})
+\frac{i}{2}r_{|\omega|,\v{q}}v_{\omega,\v{q}}^2\right]+\ord{T^0}
\ee
where $T=t_f-t_i$ and $V$ is the three dimensional volume. The second term term stands for the contribution of the 
CTP boundary conditions and will be ignored in the limit $T\to\infty$, needed to use the propagator of Appendix \ref{ctpprop}.

It is well known that the dependence of the single time-axis path integral on the final coordinates reproduces the 
wave function(al) of the system. This makes it natural to consider the integrand of the path integral for a
given trajectory as the contribution of the trajectory to the transition amplitude. In a similar manner we can
interpret the integrand in Eq. \eq{efth} as the contribution of a pair of trajectories to the reduced density
matrix of the photon field.  The $\ord{T}$ part of the effective action gives the contribution
\be\label{demt}
\Delta\rho\left(u+\frac{v}{2},u-\frac{v}{2}\right)=e^{-\frac{b}{2}v^2+iauv},
\ee
for a given mode to the reduced density matrix up to a field independent normalization constant where 
$a,b=\ord{V}$. This distribution yields finite moments $\la u^n\ra$ for large but finite volume and vanishing 
average $\la v^n\ra=0$ . The mixed moments arising in the higher order of the 
perturbation expansion can be non-vanishing due to the strong correlation between $u$ and $v$ realized by the
imaginary part of the exponent, $\la u^{2n}\ra/\la u^nv^n\ra=\ord{(b/a)^n}$. Thus we take
\be
C=\frac{b}{a}\approx\frac{\bar u}{\overline{v}},
\ee
the approximate inverse ratio of the off-diagonal and the quantum fluctuations, as a measure of the 
strength of the decoherence. The comparison of the effective action \eq{effevfsp} with 
the logarithm of the density matrix \eq{demt} shows that this quantity what we shall call classicality is
\be\label{class}
C_{\omega,\v{q}}=\frac{r_{|\omega|,\v{q}}}{\v{q}^2-L_{\omega,\v{q}}}
\ee
for the Coulomb field. 

The decoherence of the Coulomb field modes, the suppression of the off-diagonal matrix elements of the 
reduced density matrix arises from the small overlap of the electron-hole polarization clouds 
for sufficiently different $u^+_{\omega,\v{q}}$ and $u^-_{\omega,\v{q}}$. The CTP formalism is actually used 
here to calculate the overlap of dressed photon states in the Fock-space, corresponding to the Coulomb field 
$u^+$ and $u^-$ rather than expectation values.

We now turn to the other signature of the quantum-classical crossover, the issue of \tib. The 'mass-shell' condition is the 
vanishing of the denominator of the classicality. Thus modes with finite classicality correspond to virtual 
quasi-particles and give no contributions to the asymptotic states according to the reduction formulas. What 
we need for classical behavior is the finite life-time due to the diffusion of the state vector into other sectors 
of the Fock-space rather than due to the virtuality of the state. Thus classical behavior is expected for $C=\infty$ 
only and $1/C$ can be considered as a distance of the mode in question from the classical domain.

Quasi-particles are identified by the pairs $(\Omega_\v{q},\v{q})$ satisfying the 'mass-shell' condition,
$\v{q}^2-L_{\Omega_\v{q},\v{q}}=0$ and one associates a harmonic oscillators to each quasi-particle in the canonical
quantization of the photon field. The complication arising from the negative norm of the temporal, Coulomb photons
is not essential from the point of view of our problem. We shall
see below, in section \ref{points} that there are two normal modes and thus two harmonic oscillators corresponding 
to each sufficiently small wave vector $\v{q}$. 

In a qualitative, heuristic treatment the classicality appears as the product of two factors, 
\be\label{cltib}
C=\frac{E_\mr{diff}}{E_\mr{en}}\cdot R.
\ee
The first factor is the ratio of the efficiency of the diffusion of the photon state into the particle-hole sector and the 
efficiency of the energy exchange within the photon sector for each mode $(\omega,\v{k})$ 
which contributes to the quasi-particle $(\Omega_\v{k},\v{k})$, ie. for $\omega\approx\Omega_\v{k}$. 
The efficiencies, $E_\mr{diff}$ and $E_\mr{en}$ are estimated by  the velocity square of the decrease of the 
amplitude due to the diffusion of the photon state into the particle-hole sector and that of the energy exchange 
within the photon sector, respectively, $E_\mr{diff}\approx(u_m/\tau)^2$,  $u_m$ being the amplitude of the 
oscillating mode and $\tau$ denotes the life-time  and $E_\mr{en}\approx(u_m\Omega)^2$. 
The other factor, $R$, represents the correction taking into account that the finite life-time is more
important for the quantum-classical crossover if the mode is closer to being 'on-shell'. A convenient
distance of a mode $(\omega,\v{q})$ form the 'mass-shell' condition is $|\omega^2-\Omega^2_\v{q}|$ thus we take 
$R=\Omega^2_\v{q}/|\omega^2-\Omega^2_\v{q}|$ and \eq{cltib} agrees with \eq{class}.
In short, the classicality $\Im(D^{++})^{-1}/\Re(D^{++})^{-1}$ balances the role the 'mass-shell' condition and 
finite life-time plays in the perturbation expansion. The first factor represents the strength of \tib~and the second 
one measures the distance from mass-shell. 

Note that the same measure for the strength for decoherence and for \tib~emerges due to the 
structure of the CTP self-energy on the right hand side of Eq. \eq{hdinv} which relates the off-diagonal 
and diagonal matrix elements of the inverse propagator.

\section{Generation of pointer states by blocking}\label{points}
We shall approach the quantum-classical crossover by starting with a system containing the short distance modes only 
and by turning on gradually the longer length modes. In other word, we introduce an infrared cut-off in the
effective theory  \eq{efth}, $k_\mr{IR}$ or $\omega_\mr{IR}$ on the eigenvalues of the covariant 
momentum or energy operator for the allowed modes space, respectively. This is the implementation of the
strategy of the renormalization group, applied to the density matrix \cite{ed}. 
Let us consider a sequence of partitions of the system plus environment 
linear space, ${\cal H}={\cal H}^{(1)}_\Lambda\otimes{\cal H}^{(2)}_\Lambda$, parametrized by 
a gauge invariant separation scale $\Lambda=k_\mr{IR}$ or $\omega_\mr{IR}$. This scale can be introduced 
in manner that ${\cal H}^{(1)}_0$ is the linear space of 
the system $S$ and ${\cal H}^{(1)}_\Lambda\subset{\cal H}^{(1)}_{\Lambda'}$ for $\Lambda<\Lambda'$. 
The renormalization group consists of the construction of the sequence of effective theories and reduced density 
matrices $\rho_\Lambda$ for the subspace ${\cal H}^{(1)}_\Lambda$. The transformation 
$\rho_\Lambda\to\rho_{\Lambda'}$, $\Lambda'<\Lambda$, is called blocking. The question we are interested
is the qualitative way the classical physics is approached as $\Lambda$ decreased.

\begin{figure}
\includegraphics[height=3.cm,width=4.cm]{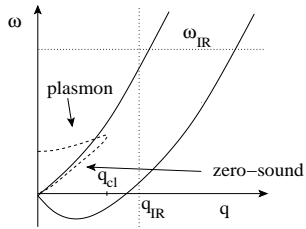}
\caption{The frequency-wave vector plane of Coulomb photons.}\label{imcar}
\end{figure}

Let us first follow blocking in three-space where the spatial resolution length of the effective theory
$1/k_\mr{IR}$ is increased to $1/(k_\mr{IR}-\Delta k_\mr{IR})$ by including modes 
$k_\mr{IR}-\Delta k_\mr{IR}<k<k_\mr{IR}$ in the effective theory \eq{ceffa}. We follow this procedure
qualitatively on the frequency-wave vector plane of Fig. \ref{imcar} in the simplest non-trivial approximation 
where the quadratic part of the effective action is kept fixed. As $k_\mr{IR}$ is lowered modes with 
wave-numbers around  $k_\mr{IR}$, the vertical dotted line appear in the effective theory. 

\begin{figure}
\includegraphics[height=4.cm,width=5.cm]{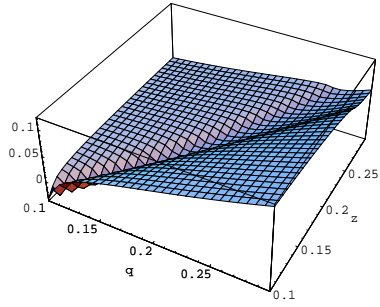}
\caption{$\v{q}^2-L_{|\omega|,\v{q}}$ plotted on the segment $0.1<q,z<0.3$ of the frequency-wave 
vector plane for $r_s=0.05$. The function has a valley stretching from the origin along the line $z=q$. The 
function is increasing in the valley as we move away from the origin and the minimum of the valley
reaches zero at $z\approx q=0.1875$.}\label{denomcl}
\end{figure}

\begin{figure}
\includegraphics[height=4.cm,width=5.cm]{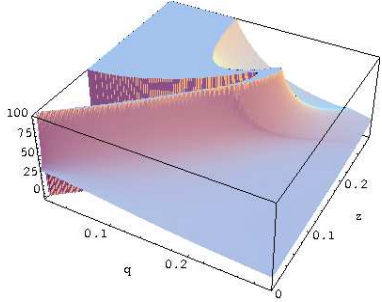}
\caption{$1/(\v{q}^2-L_{|\omega|,\v{q}})$ as the function of $z$ and $q$ for $r_s=0.05$. The function
diverges along the zero-sound/plasmon line but its value is cut at 100 for easier visualization.}\label{ree}
\end{figure}

\begin{figure}
\includegraphics[height=4.cm,width=5.cm]{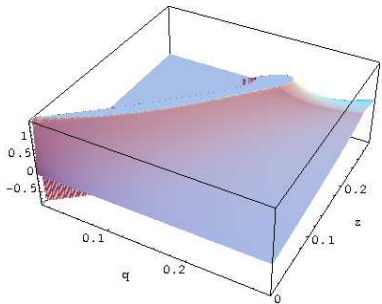}
\caption{The classicality, Eq. \eq{class}, as the function of $z$ and $q$ for $r_s=0.05$. The 
shape of this function is qualitatively similar to that shown in Fig. \ref{ree} except the part with vanishing 
$r_{\omega,\v{q}}$ is replaced by zero. The value of the function is cut at 1.5 for easier visualization.}\label{cle}
\end{figure}

What do we know about the modes of the Coulomb photon? These modes are non-propagating
on the tree-level due to the absence of the frequency dependence of the free propagator, \eq{phtpr}.
The dressed inverse propagator $\hD^{-1}$ displays frequency dependence and the collective modes, 
defined by the 'mass-shell' condition $\Re(D^{-1})^{++}_{\omega,\v{k}}=\v{q}^2-L_{\omega,\v{q}}=0$, 
are propagating. They are located on the dashed line in Fig. \ref{imcar}. The location of this curve can be seen 
clearly from the plot of $\v{q}^2-L_{\omega,\v{q}}$ in Fig. \ref{denomcl}. The tilted and the horizontal parts 
of the dashed line of Fig. \ref{imcar} are natural to identify with the zero-sound and plasmon modes, respectively. 
The corresponding factor of the classicality, $1/(\v{q}^2-L_{|\omega|,\v{q}})$, is depicted in Fig. \ref{ree}. 
Finally, the complete ratio \eq{class} for the classicality is shown in Fig. \ref{cle}.

 Another information conveyed by Fig. \ref{imcar} is that the modes between the solid line parabolas can 
decay into a real particle-hole pair. These modes acquire non-vanishing imaginary part in their inverse 
causal propagator calculated in the one-loop  approximation,  $\Im(D^{-1})^{++}_{\omega,\v{k}}=r_{|\omega|,\v{q}}>0$
and generate a Gaussian suppression of the reduced density matrix in the off-diagonal direction. 

The theory where we cut out the modes $k<k_\mr{IR}\gg k_F$ has no classical regime, having no propagating modes. 
The situation changes drastically when the vertical dotted line erected at $q=k_\mr{IR}$ approaches the 
rightmost position of the collective modes curve what happens at $k_{IR}=q_\mr{cl}$. From now on the further
decrease of the infrared cut-off brings in modes with diverging classicality and they render the effective theory classical.
When the energy rather than the wave vector of the modes is bounded from below by $\omega_{IR}$ then we find 
quantum behavior when the horizontal dotted line of Fig. \ref{imcar}, drawn at $\omega=\omega_{IR}$ is above
the uppermost point of the collective mode curve at $\omega_\mr{cl}$. Classical physics is reached when 
$\omega_{IR}=\omega_\mr{cl}$.  This scheme is closer to the actual experiments aiming at the dynamics of 
the collective modes where some energy is injected and becomes dissipated over different length scales.

What happens when higher loop contributions are taken into account in the effective theory?
Both the numerator and the denominator of the classicality change. The change of the denominator
may shift the zero-sound/plasmon line considerably. This modification is small at 
high density where loop-expansion is reliable. The higher order corrections are supposed to
render the numerator non-vanishing over the whole $(\omega,\v{q})$ plane by means of the decay
of the Coulomb photon into into multiple particle-hole pairs in the absence of gap in the fermion excitation 
spectrum. This introduces diverging classicality along the whole plasmon line. 

It is worthwhile noting that the classicality as given in Eq. \eq{class} displays the competition of tendencies,
the classical and quantum one, represented by the numerator and denominator, respectively.
The numerator is regular in the vicinity of the collective mode curve of Fig.  \ref{imcar}. It is the 
fast changing quantum features traced by the denominator which drive the crossover to classical
behavior. 

\begin{figure}
\includegraphics[height=6.cm,width=5cm,angle=270]{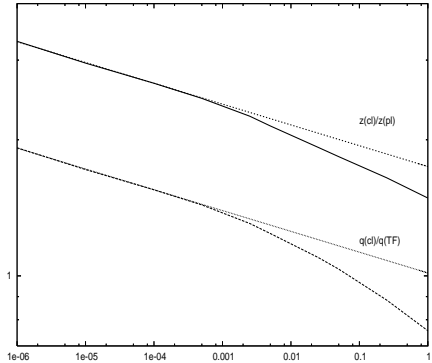}
\caption{The ratios $\omega_\mr{cl}/\omega_\mr{pl}$ and $k_\mr{cl}/k_\mr{TF}$ as the functions of $r_s$.
The straight lines correspond to the fit \eq{fitde} and \eq{fitdk}.}\label{dens}
\end{figure}

It is interesting to follow the way the scales of the quantum-classical crossover change with the density.
There are two characteristic plasma scales. One is the plasmon oscillation frequency for vanishing wave vector, 
\be
z_\mr{pl}=\sqrt{\frac{4\kappa r_s}{3\pi}},
\ee
and the other is Thomas-Fermi screening length defined at vanishing frequency,
\be
q^2_\mr{TF}=\frac{4\kappa r_s}{\pi}k_F^2.
\ee
Both are given here in the one-loop approximation. It was found that the wave vector $q_\mr{cl}$ 
and frequency $\omega_\mr{cl}$ of the mode which triggers the quantum-classical 
transition during the blocking is close to the plasmon frequency and the Thomas-Fermi screening
scale, respectively. The plot of the ratios $z_\mr{cl}/z_\mr{pl}$ and $q_\mr{cl}/q_\mr{TF}$ as the functions
of the density are shown in Fig. \ref{dens}, together with the fitted form
\be\label{fitde}
z_\mr{cl}\approx1.745\cdot r_s^{-0.046}z_\mr{pl}=0.822\cdot r_s^{0.495},
\ee
and
\be\label{fitdk}
q_\mr{cl}\approx1.015\cdot r_s^{-0.046}q_\mr{TF}=0.822\cdot r_s^{0.495}
\ee
which are valid for high density, small $r_s$. 

The relation $z_\mr{cl}\approx q_\mr{cl}$ is not surprising, 
it is the obvious result of the position of the zeros of $\Re(D^{-1})^{++}$ in the vicinity of $\omega=\v{k}=0$. 
The non-trivial result is the approximate agreement between $q_\mr{cl}$ and $q_\mr{TF}$.
In fact, the former locates the root of $\Re(D^{-1})^{++}_{\omega,\v{q}}$ at $\omega\approx\omega_\mr{cl}\not=0$ 
while the latter is related to the value of the $\Re(D^{-1})^{++}_{\omega,\v{q}}$ for $\omega=0$ and $\v{q}\to0$.
Their agreement, that the classical behavior is reached at the screening length, shows the importance of the
screening cloud in the classical crossover from the point of view of the charge dynamics. But this is valid in the
absence of the intrinsic scale of the photon dynamics only. In fact, let us introduce a finite mass $M^2$ in the photon
propagator \eq{phtpr} which makes the static potential Yukawa-type. This raises 
$\v{q}^2-L_{|\omega|,\v{q}}$, shown in Fig. \ref{denomcl} by $M^2$ and shrinks the collective mode curve of
Fig. \ref{imcar} towards the origin. The screening is reduced to a simple renormalization of the mass,
$M^2\to M^2+\delta M^2$,
and the agreement $q_\mr{cl}\approx q_\mr{TF}$ is lost, the massive photon and the charge dynamics
now have two distinct scales. Our approximate relation, $q_\mr{cl}\approx q_\mr{TF}$, observed for mas-less
photons indicates that the renormalization of the mass square of the photon due to the environment,
$\delta M^2=q^2_\mr{TF}$, agrees with the crossover scale, $q^2_\mr{cl}\approx\delta M^2$ generated by the
same environment in the one-loop level. Naturally these relations receive corrections from higher orders of the loop-expansion.
It is interesting to observe that no collective modes and classical behavior is reached for sufficiently large $M^2$.

\section{Conclusions}\label{concls}
The contributions to decoherence and the dynamical breakdown of the time reversal invariance 
of the Coulomb field dynamics were calculated in the framework of the CTP formalism in this paper. 
A quantity to measure the contributions of the plane wave modes of the Coulomb field
to decoherence and \tib, the classicality was proposed. It was found in the one-loop approximation
that the collective modes, corresponding to the zero-sound and plasmon excitations have diverging classicality
and render the system classical. This result has already been expected, our calculation shows that the 
relation between decoherence and \tib~ is the result
of the structure of the CTP propagators and establishes the classicality of the collective modes in
a systematic manner as the result of the gap-less excitation spectrum of the environment. 
The length scale of the quantum-classical crossover is found to be
close to the Thomas-Fermi screening length. The dressed quasi-particles, corresponding to the
collective modes of the plasma and being controlled by the Coulomb field are formed by minimizing their
residual interactions. This minimization makes them dynamically optimized pointer states in every order of the
loop-expansion.

Another, indirect lesson of this calculation is the suitability of quantum field theoretical
models to the problem of measurement theory. The advantages are (i) the possibility to consider
a large number of degrees of freedom, (ii) the way to test dynamical symmetry breaking, such as the
loss of invariance under time inversion and (iii) the accessibility of the reduced density matrix
in the CTP formalism. Finally, the strategy of the renormalization group seems to be well
suited to address the quantum-classical transition.

\acknowledgments
It is pleasure to thank Janos Hajdu for stimulating discussions.

\appendix
\section{CTP propagator for non-relativistic particles}\label{ctpprop}
Let us consider free non-relativistic particles with exchange statistics $\xi=\pm$. Their propagator will be
identified by means of the generator functional
\be
e^{\ih W[\hj,\bar{\hj}]}=\Tr T[e^{-\ih\int_{t_i}^{t_f}dt'[H(t')-\bar j_+(t')\psi_+(t')-\psid_+(t')j_+(t')]}]\rho_i
\bar Te^{\ih\int_{t_i}^{t_f}dt'[H(t')+\psi_-(t')\bar j_-(t')+j_-(t')\psid_-(t')]}].
\ee

\subsection{Path integral}
The generator functional, written in the path integral representation reads
\be
e^{\ih W[\hj,\bar{\hj}]}\int D[\hsi]D[\hsid]e^{\ih\hsid\cdot\hG^{-1}\cdot\hsi+\ih\bar{\hj}\cdot\hsi+\ih\hsid\cdot\hj}
\ee
and it yields
\be
W[\hj,\bar{\hj}]=-\bar{\hj}\cdot\hG\cdot\hj.
\ee
The second function derivative of the generator functional gives
\bea\label{nrpprop}
\xi i\hbar G_{xy}&=&i\fdd{W[\hj,\bar{\hj}]}{i\bar j^+_x}{ij^+_y}
=\xi\sum_n\la0|n\ra\la n|T[\psi_x\psid_y]|0\ra=\xi\la0|T[\psi_x\psid_y]|0\ra\nn
\xi i\hbar G^{--}_{xy}&=&i\fdd{W[\hj,\bar{\hj}]}{i\bar j^-_a}{ij^-_b}
=\xi\sum_n\la0|\bar T[\psi_x\psid_y]|n\ra\la n|0\ra=\xi\la0|\bar T[\psi_x\psid_y]|0\ra=\xi\la0|T[\psi_y\psid_x]|0\ra^*\nn
\xi i\hbar G^{+-}_{xy}&=&i\fdd{W[\hj,\bar{\hj}]}{i\bar j^+_x}{ij^-_y}
=\xi\sum_n\la0|\psid_y|n\ra\la n|\psi_x|0\ra=\xi\la0|\psid_yU^\dagger(t_f,t_y)U(t_f,t_x)\psi_x|0\ra\nn
\xi i\hbar G^{-+}_{xy}&=&i\fdd{W[\hj,\bar{\hj}]}{i\bar j^-_x}{ij^+_y}
=\sum_n\la0|\psi_x|n\ra\la n|\psid_y|0\ra=\la0|\psi_x\psid_y|0\ra
\eea
and we find in this manner
\be
i\hbar\begin{pmatrix}G&G^{+-}\cr G^{-+}&G^{--}\end{pmatrix}_{xy}
=\begin{pmatrix}\la0|T[\psi_x\psid_y]|0\ra&\la0|\psid_y\psi_x]|0\ra\cr
\xi\la0|\psi_x\psid_y|0\ra&\la0|T[\psi_y\psid_x]|0\ra^*\end{pmatrix}
\ee

The CTP propagator, a 2x2 matrix contains three independent functions only due to the identity
\be
T[\psi\psid]+\bar T[\psi\psid]=\xi\psid\psi+\psi\psid
\ee
implying
\be
G+G^{--}=\xi G^{+-}+\xi G^{-+}.
\ee

\subsection{Operator formalism}
The actual form of the free propagator is easier to obtain in the operator formalism. We assume discrete
momentum spectrum for the sake of simplicity where the canonical creation and destruction operators
satisfy the commutation relations 
\be
[b_\v{p},b^\dagger_{\v{p}'}]_\xi=V\delta_{\v{p},\v{p}'}
\ee
and
\be
[b_\v{p},b_{\v{p}'}]_\xi=0,
\ee
and the quantum field is written as
\be
\psi_x=\frac1V\sum_\v{k}b_\v{k}e^{-ikx}.
\ee
The initial density matrix is assumed to be of the form
\be
\rho_i=\frac{1}{\Tr[e^{-\frac{\beta}{V}\sum_\v{k}\epsilon_\v{k}b^\dagger_\v{p}b_\v{p}}]}
e^{-\frac{\beta}{V}\sum_\v{k}\epsilon_\v{k}b^\dagger_\v{p}b_\v{p}}
\ee
where $\epsilon_\v{k}=E_\v{k}-\mu$.

We calculate $G^{-+}$ first. The starting point is the last equation of Eq. \eq{nrpprop},
\bea
i\hbar G^{-+}_{xx'}\Tr\rho&=&\xi\Tr\rho\psi_x\psid_{x'}\nn
&=&\frac{\xi}{V^2}\sum_\v{k}\Tr e^{-\frac{\beta}{V}\sum_\v{k}\epsilon_\v{k}b^\dagger_\v{p}b_\v{p}}b_\v{k}b^\dagger_\v{k}e^{-ik(x-x')}\nn
&=&\xi\Tr\rho\frac1V\sum_\v{k}e^{-ik(x-x')}\frac{\sum_ne^{-\beta n\epsilon_\v{k}}(1+\xi n)}{\sum_ne^{-\beta n\epsilon_\v{k}}}.
\eea
Its Fourier transform,
\be
i\hbar G^{-+}_k=\int_xe^{ik(x-x')}i\hbar G^{-+}_{xx'},
\ee
can be written as
\be
i\hbar G^{-+}_k=2\pi\delta(k^0-\epsilon_\v{k})(\xi+n^\xi_\v{k}),
\ee
where 
\be
n^\xi_\v{k}=\frac{1}{e^{\beta\epsilon_\v{k}}-\xi}.
\ee

One finds in a similar manner
\bea
i\hbar G^{+-}_{xx'}\Tr\rho&=&\Tr\rho\psid_{x'}\psi_x\nn
&=&\frac{1}{V^2}\sum_\v{k}\Tr e^{-\frac{\beta}{V}\sum_\v{k}\epsilon_\v{k}b^\dagger_\v{p}b_\v{p}}b^\dagger_\v{k}b_\v{k}e^{-ik(x-x')}\nn
&=&\Tr\rho\frac1V\sum_\v{k}e^{-ik(x-x')}\frac{\sum_ne^{-\beta n\epsilon_\v{k}}n}{\sum_ne^{-\beta n\epsilon_\v{k}}}
\eea
and
\be
i\hbar G^{+-}_k=2\pi\delta(k^0-\epsilon_\v{k})n^\xi_\v{k}.
\ee

The component $G=G^{++}$ is the causal propagator,
\be
G_{xx'}=\xi[\Theta(t-t')G^{-+}_{xx'}+\Theta(t'-t)G^{+-}_{xx'}]
\ee
which can be written as
\be
i\hbar G_{xx'}=-\frac{\xi}{i}\int_\omega\frac{e^{-i\omega(t-t')}}{\omega+i\epsilon}
\int_\v{k}(\xi+n^\xi_\v{k})e^{-i(\epsilon_\v{k}-i\gamma)(t-t')+i\v{k}(\v{x}-\v{x'})}
-\frac{\xi}{i}\int_\omega\frac{e^{i\omega(t-t')}}{\omega+i\epsilon}
\int_\v{k}n^\xi_\v{k}e^{-i(\epsilon_\v{k}+i\gamma)(t-t')+i\v{k}(\v{x}-\v{x'})}.
\ee
This form gives after trivial rearrangements
\be
i\hbar G_{xx'}=i\int_k\frac{e^{-ik(x-x')}}{\omega-\epsilon_\v{k}+i\epsilon}
+2\pi\int_k\delta(\omega-\epsilon_\v{k})\xi n^\xi_\v{k}e^{-ik(x-x')}
\ee
and
\be
\hbar G_k=\frac{1}{k^0-\epsilon_\v{k}+i\epsilon}-2i\pi\delta(k^0-\epsilon_\v{k})\xi n^\xi_\v{k}.
\ee

The complete CTP propagator is finally
\be
\begin{pmatrix}G&G^{+-}\cr G^{-+}&G^{--}\end{pmatrix}_k
=\begin{pmatrix}\frac{1}{k^0-\omega_\v{k}+i\epsilon}-i2\pi\delta(k^0-\epsilon_\v{k})\xi n^\xi_\v{k}&
-i2\pi\delta(k^0-\epsilon_\v{k})n^\xi_\v{k}\cr
-i2\pi\delta(k^0-\epsilon_\v{k})(\xi+n^\xi_\v{k})&
\frac{1}{\omega_\v{k}-k^0+i\epsilon}-i2\pi\delta(k^0-\epsilon_\v{k})\xi n^\xi_\v{k}\end{pmatrix}.
\ee

\section{Self energy}\label{tgap}
The Fourier transform of the self energy of the Coulomb field, \eq{cself},
\be
\hat\Sigma^{\sigma\sigma'}_{xx'}=\int_q\hat\Sigma^{\sigma\sigma'}_qe^{-i(x-x')q}
\ee
is
\bea
\hat\Sigma^{\sigma\sigma'}_q&=&-2ie^2\hbar\sigma\sigma'\int_pG^{\sigma\sigma'}_{p+q}G^{\sigma'\sigma}_pe^{4i\eta p^0}\nn
&=&-2ie^2\hbar\sigma\sigma'\int_k\begin{pmatrix}\frac{1}{k^0+q^0-\epsilon_{\v{k}+\v{q}}+i\epsilon}
+i2\pi\delta(k^0+q^0-\epsilon_{\v{k}+\v{q}})n_{\v{k}+\v{q}}&
-i2\pi\delta(k^0+q^0-\epsilon_{\v{k}+\v{q}})n_{\v{k}+\v{q}}\cr 
i2\pi\delta(k^0+q^0-\epsilon_{\v{k}+\v{q}})(1-n_{\v{k}+\v{q}})&
\frac{1}{\epsilon_{\v{k}-\v{q}}-k^0-q^0+i\epsilon}
+i2\pi\delta(k^0+q^0-\epsilon_{\v{k}+\v{q}})n_{\v{k}-\v{q}}\end{pmatrix}^{\sigma\sigma'}\nn
&&\times\begin{pmatrix}\frac{1}{k^0-\epsilon_\v{k}+i\epsilon}+i2\pi\delta(k^0-\epsilon_\v{k})n_\v{k}&
-i2\pi\delta(k^0-\epsilon_\v{k})n_\v{k}\cr
i2\pi\delta(k^0-\epsilon_\v{k})(1-n_\v{k})&\frac{1}{\epsilon_\v{k}-k^0+i\epsilon}
+i2\pi\delta(k^0-\epsilon_\v{k})n_\v{k}\end{pmatrix}^{\sigma'\sigma}e^{4i\eta p^0},
\eea
where $\eta=0^+$. The $++$ matrix element,
\be
\hat\Sigma^{++}_q=-2ie^2\hbar\int_k\left(\frac{1}{k^0+q^0-\epsilon_{\v{k}+\v{q}}+i\epsilon}
+i2\pi\delta(k^0+q^0-\epsilon_{\v{k}+\v{q}})n_{\v{k}+\v{q}}\right)
\left(\frac{1}{k^0-\epsilon_\v{k}+i\epsilon}+i2\pi\delta(k^0-\epsilon_\v{k})n_\v{k}\right)
\ee
can be written as a sum, $\hat\Sigma^{++}=\hat\Sigma^{(0)++}+\hat\Sigma^{(1)++}$, where
\bea
\hat\Sigma^{(0)++}_q&=&-2ie^2\hbar\int_k\frac{1}{(k^0+q^0-\epsilon_{\v{k}+\v{q}}+i\epsilon)(k^0-\epsilon_\v{k}+i\epsilon)}\nn
\hat\Sigma^{(e)++}_q&=&2e^2\hbar\int_\v{k}\left[\frac{n_\v{k}}{\epsilon_\v{k}+q^0-\epsilon_{\v{k}+\v{q}}+i\epsilon}
+\frac{n_{\v{k}+\v{q}}}{-q^0+\epsilon_{\v{k}+\v{q}}-\epsilon_\v{k}+i\epsilon}
+i2\pi\delta(\epsilon_\v{k}+q^0-\epsilon_{\v{k}+\v{q}})n_{\v{k}+\v{q}}n_\v{k}\right].
\eea
It is easy to see that the first expression, $\hat\Sigma^{(0)++}_q$ is vanishing. As of the second one, one finds
the Lindhard function
\be
L_q=\Re\hat\Sigma^{(e)++}_q=2\hbar e^2P\int_\v{k}\frac{n_\v{k}-n_{\v{k}+\v{q}}}{q^0-\epsilon_{\v{k}+\v{q}}+\epsilon_\v{k}},
\ee
where $P$ denotes the principal value integral for the real part. Straightforward integration yields Eq. \eq{lindhfn}.

The imaginary part,
\be
\Im\hat\Sigma^{(e)11}_q=-i\hbar e^2\int_\v{k}2\pi\delta(q^0-\epsilon_{\v{k}+\v{q}}+\epsilon_\v{k})(n_\v{k}+n_{\v{k}+\v{q}}-2n_{\v{k}+\v{q}}n_\v{k})
\ee
can be written as
\be
\Im\hat\Sigma^{(e)11}_{\omega,\v{q}}=-i(r_{\omega,\v{q}}+r_{-\omega,\v{q}})
\ee
where
\be
r_{\omega,\v{q}}=\hbar e^2\int_\v{k}2\pi\delta(\omega-\epsilon_{\v{k}-\v{q}}+\epsilon_\v{k})n_\v{k}(1-n_{\v{k}-\v{q}}).
\ee
The integral is the area of a part of a plan which is within the Fermi-sphere located at the origin but
outside of another Fermi-sphere, shifted by $\v{q}$ and can be carried out easily, leading for $\omega>0$ to
Eq. \eq{limag}.

The other, remaining matrix elements of the self energy are
\be
\hat\Sigma^{\pm\mp}_{\omega,\v{q}}=2\Theta(\mp\omega)r_{|\omega|,\v{q}}.
\ee


\begin{thebibliography}{99}
\bibitem{tom} S. Tomonaga, \journal{Prog. Theor. Phys.}{1}{27}{1946}.
\bibitem{schwe} J. Schwinger, V. Weisskopf, \journal{Phys. Rev.}{73}{1272A}{1948}.
\bibitem{schw} J. Schwinger, \journal{J. Math. Phys.}{2}{407}{1961};
{\em Particles and Sources}, vol. I., II., and III.,Addison-Wesley, Cambridge, Mass. 1970-73.
\bibitem{keldysh} L. V. Keldysh, \journal{Zh. Eksp. Teor. Fiz.}{47}{1515}{1964}
(\journal{Sov. Phys. JETP}{20}{1018}{1965}).
\bibitem{ed} J. Polonyi, \journal{Phys. Rev.}{D74}{065014}{2006}.
\bibitem{zeh} H. D. Zeh, journal{Found. Phys.}{1}{1970}.
\bibitem{zurek} W. H. Zurek, \journal{Phys. Rev.}{D24}{1516}{1981}.
\bibitem{pointer} W. H. Zurek, \journal{Phys. Rev.}{D26}{1862}{1982}.
\bibitem{optpoint} W. H. Zurek, S. Habib, J. P. Paz, \journal{Phys. Rev. Lett.}{70}{1187}{1993}.
\bibitem{wiva} H. M. Wisemann, J. A. Vaccaro, \journal{Phys. Rev.}{A65}{043606}{2002}.
\bibitem{dzdazu} J. Dziarmaga, A. R. Dalvit, W. H. Zurek, \journal{Phys. Rev.}{A69}{022109}{2004}.
\bibitem{dadzzu} D. A. Dalvit, J. Dziarmaga, W. H. Zurek, \journal{Phys. Rev.}{A72}{062101}{2005}.
\bibitem{elze} H.Y. Elze, in the {\em Proceedings of the 5th Rio de Janeiro International Workshop on
Relativistic Aspects of Nuclear Physics}, eds. T. Kodama et al. (World Scientific), arXiv:quant-ph/9710063.
\bibitem{dreyer} O. Dreyer, \journal{J. Phys. Conf. Ser.}{67}{012051}{2007}.
\bibitem{grkemo} B. Groisman, D. Kenigsberg, T. Mor, {\em "Quantumness" versus "Classicality" of Quantum States},
Preprint arXiv:quant-ph/0703103, (2007).
\bibitem{mapa} G. Mahajan, T. Padmanabhan {\em Particle creation, classicality and related issues in quantum field theory: 
I. Formalism and toy models}, Preprint arXiv:0708.1233, (2007).
\bibitem{dore} V.V. Dodonov, M. B. Reno, \journal{Phys. Lett.}{A308}{249}{2003}.
\bibitem{arnold} V. I. Arnold, A. Avez, {\em Ergodic Problems in Classical Mechanics}, Benjamin, New York (1968).
\bibitem{peres} M. Feingold, A. Peres, \journal{Phys. Rev.}{A34}{591}{1968}.
\bibitem{feynman} R. P. Feynman, F. L. Vernon, \journal{Ann. Phys.}{24}{118}{1963}.
\end{thebibliography}
\end{document}